# Change of Hydration Patterns upon RNA Melting Probed by Excitations of Phosphate Backbone Vibrations

Achintya Kundu, Jakob Schauss, Benjamin P. Fingerhut, and Thomas Elsaesser[*]

*Max-Born-Institut für Nichtlineare Optik und Kurzzeitspektroskopie,*

*Berlin 12489, Germany*

[*]Corresponding author: elsasser@mbi-berlin.de, phone +49 30 6392 1400, fax +49 30 6392 1409

**Abstract**. The water hydration shell has decisive impact on the structural and functional properties of RNA. Changes of RNA structure upon melting and in biochemical processes are accompanied by a change of hydration patterns, a process which is barely characterized. To discern hydration geometries around the backbone phosphate groups of an RNA double helix at the molecular level, we combine two-dimensional infrared spectroscopy of phosphate vibrations with theoretical simulations. There are three distinct coexisting hydration motifs of the RNA A-helix, an ordered chain-like arrangement of water molecules with links between neighboring phosphate groups, separate local hydration shells of up to six water molecules, and hydrated phosphate/counterion contact pairs. RNA disordering upon melting is connected with a transition from predominant ordered water structures to local hydration shells around phosphate units. Structural fluctuations are dominated by librational water motions occurring on a 300 fs time scale, without exchange between hydration motifs.

# 1. Introduction

The macromolecular structure of RNA plays a key role for its function in biochemical processes.[1] Folded secondary structures of RNA are stabilized by interactions with their hydration shell and (counter)ions embedded in the aqueous environment.[2,3] Key interactions are local hydrogen bonds and longer-range Coulomb forces.[4-6] Most important in the folding of RNA structures is the overcoming of the strong electric repulsion between negatively charged phosphate groups in the RNA backbone mediated by (counter)ions and dipolar water molecules. Changes of secondary structure, e.g., by RNA folding, during RNA translation in cells or melting of RNA structures at elevated temperature are inevitably connected with changes of hydration geometries that have not been understood so far.

Time-averaged structures of hydrated RNA have been studied under quasi-equilibrium conditions by x-ray diffraction from crystallized samples and by molecular dynamics simulations.[7-9] At ambient temperature (~300 K), fully hydrated double-stranded RNA typically forms an A-helix with adenine-uracil and guanine-cytosine pairs in a hydrogen-bonded Watson-Crick geometry. In the A-helix, the free $(PO_2)^-$ oxygens of the backbone point into the grooves at the RNA surface. The structure of the first water layer is mainly determined by the steric boundary conditions and charge distribution at this surface. Adjacent $(PO_2)^-$ units are bridged by individual water molecules while the ribose 2'-OH groups induce chain-like arrangements of hydrogen bonded water molecules in the minor groove. Compared to the DNA B-helix,[10] the water structure at the RNA surface is more ordered.[7,11]

Hydrated DNA and RNA display fluctuations of equilibrium structure and vibrational dynamics in the femtosecond time domain, complemented by a breaking and reformation of water hydrogen bonds in the picosecond time range.[4,12-14] The ionic and polar groups of the backbone, in particular the phosphate groups, are subject to fluctuating electric fields from the hydration shell and vice versa exert electric forces on the dipolar water molecules.[15] Recently, we have applied two-dimensional infrared (2D-IR) spectroscopy of backbone vibrations to unravel hydration dynamics around the B-helix of DNA and the A-helix of RNA at ambient temperatures around 300 K.[11,16-18] The analysis of the 2D line shapes reveals fast fluctuations of the solvating water shell on a 300 fs time scale, while structural disorder of the helix-water interface translates into an inhomogeneous distribution of vibrational transition frequencies, persisting for at least tens of picoseconds. Ab-initio calculations and molecular dynamics simulations suggest short-range electric interactions with time-averaged field amplitudes of ~90 MV/cm at the RNA/DNA surfaces and fluctuation amplitudes of ~25 MV/cm.[18]

Here, we combine femtosecond 2D-IR spectroscopy and theoretical simulations to explore changes of the molecular hydration pattern upon RNA melting, a prototypical change of secondary structure. The asymmetric $(PO_2)^-$ stretching vibrations $\nu_{AS}(PO_2)^-$ of the backbone of double-stranded RNA oligomers serve as dynamical noninvasive probes which are sensitive to both local hydrogen bonds with the free $(PO_2)^-$ oxygens and to electric fields due to the high electric polarizability of the phosphate groups. The results for double-helical structures are benchmarked by data for single-stranded RNA and double-stranded DNA.

## 2. Methods

In the experiments, double-stranded RNA oligomers (dsRNA) containing 23 alternating adenine-uracil (A-U) base pairs in an A-helix geometry were dissolved in ultrapure $H_2O$. As benchmark systems, we studied double-stranded DNA oligomers with 23 alternating adenine-thymine (A-T) base pairs and single-stranded RNA with 23 A bases (ssRNA). In the latter, stacking of bases is possible while base pairing is absent. All samples contain $K^+$ counterions, the RNA and DNA concentrations are in the millimolar range and summarized in Table S1 of the Supporting Information (SI). The sample temperature $T_S$ was varied in a range from 295 to 355 K in order to induce changes of the RNA structures.

For an analysis of the RNA structure changes upon heating, both electronic absorption and circular dichroism spectra were measured over a wide temperature range. Heating the RNA results in a de-stacking of bases, a disordering of the molecular structures, and a concomitant increase of electronic absorption.[19] In parallel, a thermally induced change of chirality of the molecular structure modifies the circular dichroism (CD).[20] Ultraviolet absorption spectra of dsRNA and ssRNA at different sample temperatures $T_S$ are summarized in Fig. 1. With increasing temperature, the peak absorbance of dsRNA at 260 nm (inset) shows a step-like sigmoid increase from which one estimates a transition temperature to a disordered state of $T_m$ = 320±5 K (47±5 °C). The ssRNA structure containing adenine bases only displays a gradual increase of ultraviolet absorption without a clear transition point to a disordered structure. This behavior is in agreement with previous studies of adenine-rich single-stranded RNA and has been assigned to a transition from an A-helix structure of the single strand to a disordered structure by de-stacking of the adenine bases.[21,22]

CD spectra of dsRNA and ssRNA for different sample temperatures $T_S$ are presented in Fig. 2. The shape of the CD spectrum of dsRNA at 298 K exhibits a maximum at 257 nm, a dip at 245 nm, and a negative signal at 214 nm, features which are typical for an A-form double helix. The disordering of this structure with increasing temperature is evident from the gradual red-shift and loss of intensity of the positive band. The temperature dependent red-shift (inset) suggests a disordering temperature around 320 K, in agreement with the ultraviolet absorption spectra in Fig. 1. The CD spectra of both dsRNA and ssRNA are in line with literature data.[20,23,24]

Linear and femtosecond nonlinear infrared spectroscopy of the asymmetric phosphate stretching vibrations $\nu_{AS}(PO_2)^-$ was applied for mapping the impact of RNA melting on hydration geometries and their dynamics. Experimental details are explained in the SI. The 2D-IR spectra were measured in 3-pulse photon-echo experiments with heterodyne detection of the nonlinear signal. Details of this setup have been presented elsewhere.[16] The 2D-IR spectra were analyzed by calculations of the third-order nonlinear vibrational response within the framework of density matrix theory, including a simulation of 2D line shapes based on a Kubo approach for the frequency fluctuation correlation function of the system. Details of this analysis and calculated 2D-IR spectra are presented in the SI.

Molecular Dynamics (MD) simulations were performed as described in detail in Ref. (11). In brief, the initial model structure of the AU 23-mer RNA double strand was generated with the nucleic acid builder (NAB) of the AmberTools program suite as canonical Watson-Crick RNA double helix (A-form, sequence: 5'-UUAUAUAUAUAUAUAUAUAUAUU-3' and its complementing strand). The model structure was placed in a truncated octahedral solvation box with a 10.0 Å buffer region. The TIP3P water model was employed with random replacement of water molecules by 44 sodium counterions for charge neutrality. MD simulations were performed with AMBER 14 [25] employing the *ff99bsc0* force field with $\chi_{OL3}$ corrections[26] as recommended for RNA. Single valent sodium ion parameters for the TIP3P water model were taken from Refs. (27,28). Long time MD simulations (1.2 µs) were performed with the PMEMD program and the GPU accelerated PMEMD.CUDA program (Tesla K80),[29,30] further technical details are given in the SI.

Vibrational frequencies of $\nu_{AS}(PO_2)^-$ were evaluated in mixed quantum mechanical / molecular mechanical (QM/MM) density functional simulations with electrostatic embedding at different positions of the $(AU)_{23}$ dsRNA helix after 160 ns of molecular dynamics equilibration (B3LYP-D functional; C, O, H, N, Na: 6-31G* basis set , P: 6-311+G* basis set). The QM/MM models for the evaluation of vibrational frequencies of the sugar phosphate backbone of RNA were built by selecting two phosphate groups together with the three bridging ribose moieties and waters in the first solvation shell of the phosphate groups as QM region (see SI for simulation details).

## 3. Results and Discussion

In Fig. 3(a), linear infrared absorption spectra of the asymmetric $(PO_2)^-$ stretching vibrations $\nu_{AS}(PO_2)^-$ of dsRNA oligomers are presented for a wide range of sample temperatures $T_S$. The bands with maxima at 1220 and 1245 cm$^{-1}$ represent the absorption of differently hydrated (PO$_2$)$^-$ groups while the weak band at 1280 cm$^{-1}$ is due to phosphate/K$^+$ contact ion pairs as will be discussed in detail below. With increasing temperature, the $\nu_{AS}$(PO$_2$)$^-$ absorption undergoes a substantial reshaping by which the two distinct absorption peaks at 1220 and 1245 cm$^{-1}$ are replaced by a broad line shape with a maximum around 1230 cm$^{-1}$. To highlight the changes in line shape, differential absorption spectra $\Delta A = A(T_S) - A(303K)$ are plotted in Fig. 3(b) (A(T$_S$): absorbance at sample temperature T$_S$). The $\nu_{AS}$(PO$_2$)$^-$ line shapes observed at the highest temperatures of 343 and 353 K are close to the $\nu_{AS}$(PO$_2$)$^-$ absorption band of double-stranded DNA oligomers forming a B-helix (T$_S$=303 K, dashed line in Fig. 3a).

Linear absorption spectra of $\nu_{AS}$(PO$_2$)$^-$ of ssRNA are plotted in Fig. 3(c). The two components with maxima at 1220 and 1245 cm$^{-1}$ display a very similar strength at T$_S$=303 K (black line) and merge into a single broad absorption band with a maximum at 1230 cm$^{-1}$ for increasing T$_S$. In contrast to the dsRNA, there is no distinct band around 1280 cm$^{-1}$ due to contact ion pairs.

In Fig. 4, 2D-IR spectra of the $\nu_{AS}$(PO$_2$)$^-$ vibrations of (a-c) dsRNA are presented for different temperatures T$_S$ and compared to (d) DNA at T$_S$=303K. The absorptive 2D signal, i.e., the real part of the sum of the rephasing and non-rephasing photon echo signals measured at a waiting time of T=300 fs is plotted as a function of excitation frequency $\nu_1$ and detection frequency $\nu_3$. The yellow-red contours represent positive 2D-IR signals, due to ground-state bleaching and stimulated emission on the respective v=0→1 vibrational transition, while blue contours display negative 2D-IR signals due to v=1→2 absorption. The latter are red-shifted to lower detection frequencies $\nu_3$ because of the (diagonal) anharmonicity of the vibrations. The peak pattern and the 2D line shapes remain unchanged in 2D-IR spectra measured at longer waiting times up to approximately 1 ps.

In Fig. 4(a), the positive diagonal signal of dsRNA at $T_S$=303 K displays a main peak at $(\nu_1,\nu_3)$=(1242,1250) cm$^{-1}$ and two weaker components around (1220,1220) and (1283,1283) cm$^{-1}$. The different contributions are clearly discerned in the corresponding cut of the 2D-IR signal along a direction parallel to the $\nu_1=\nu_3$ diagonal line (Fig. 4e, black line). There are no cross peaks between the three contributions, as is evident from both the 2D-IR spectrum in Fig. 4(a) and cuts of this spectrum along the detection frequency axis $\nu_3$ (Fig. 4f). The absence of cross peaks points to negligible inter-mode couplings and the absence of chemical exchange between different species on the time scale of the experiment. With increasing $T_S$ (Fig. 4b,c), the maximum of the positive diagonal peak shifts to lower excitation and detection frequencies and the line shape develops into a single broad peak elongated along the diagonal $\nu_1=\nu_3$. The diagonal cuts in Fig. 4(e) (blue and red lines) exhibit a larger spectral width, corresponding to a larger inhomogeneous broadening of the main peak compared to the main peak observed at $T_S$=303 K. This behavior is in line with the corresponding changes in the linear infrared absorption spectra of Fig. 3(a). In the diagonal cuts for $T_S$=348 K, the dsRNA peak position is identical to that of the corresponding diagonal peak observed with DNA at $T_S$=303 K (dash-dotted line), with minor differences in spectral width. It should be noted that the dsRNA peak at (1283,1283) cm$^{-1}$ disappears nearly completely upon heating the sample from $T_S$=303 to 348 K.

Cuts through the main dsRNA and DNA diagonal peaks along antidiagonal directions are plotted in Fig. 4(g). Their dispersive shapes are due to the positive and negative 2D signals which partly overlap and compensate each other at detection frequencies close to the central zero crossing. There is a moderate broadening of the positive RNA signals upon heating. Population kinetics of the $\nu_{AS}(PO_2)^-$ vibration were determined in independent femtosecond pump-probe experiments (cf. SI). In RNA, the population lifetime has a value of 360±25 fs which is constant over the entire range of sample temperatures $T_S$ and close to the 340±20 fs lifetime of the vibration in DNA.[16] The 2D-IR spectra of Fig. 4 are complemented by 2D-IR spectra of ssRNA which are presented in Fig. S2 of the SI.

The 2D-IR line shapes and their changes upon heating were analyzed by numerical density matrix simulations as described in the SI. The three diagonal components in the 2D-IR spectra of dsRNA (Fig. 4a-c) are attributed to three independent, i.e., uncoupled types of excitation at 1220, 1245, and 1280 cm$^{-1}$. At T$_S$=303 K, the amplitude ratio of the 1245 cm$^{-1}$ vs 1220 cm$^{-1}$ component is approximately 2:1. The ratio decreases for higher T$_S$, leading to the changes in the diagonal cuts (Fig. 4e) and reaching a value of 1:3 at T$_S$=348 K. To account for the 2D line shapes, we apply a Kubo ansatz for the frequency fluctuation correlation function with two exponential components with decay times of 300 fs and 50 ps.[11,16] Upon heating, the amplitude of the fast component is enhanced by roughly a factor of two, reflecting a stronger thermal excitation of the underlying water librations and causing the broadening of the antidiagonal cuts (Fig. 4f).

Figure 5 presents simulation results, the infrared absorption spectra in the frequency range of the $\nu_{AS}(PO_2)^-$ vibrations obtained from QM/MM simulations at different positions of an (AU)$_{23}$ dsRNA helix (Fig. 5a), histograms of hydrogen bonds (Fig. 5b), and solvation geometries (Fig. 5c). The good agreement of the simulated and experimental infrared absorption spectra allows for a microscopic assignment of the different absorption components. Due to the particular sensitivity of $\nu_{AS}(PO_2)^-$ the differences in frequency position mirror variations in solvation structure around the phosphate groups. The component at 1220 cm$^{-1}$ (simulation: $\nu_{AS}(PO_2)^-$ = 1224.7 cm$^{-1}$, label A in Fig. 5a) is assigned to a solvation geometry with six water molecules in the first shell, i.e., the phosphate oxygen atoms are surrounded by water molecules in an approximate tetrahedral arrangement (panel A in Fig. 5c). For the component at 1245 cm$^{-1}$ (simulation: $\nu_{AS}(PO_2)^-$ = 1245.2 cm$^{-1}$, label B in Fig. 5a) we find a reduced occupancy of first solvation shell water molecules which induces a ~20-30 cm$^{-1}$ blue shift of $\nu_{AS}(PO_2)^-$. In the particular realization, a water molecule bridges two adjacent phosphate groups (panel B in Fig. 5c, shortest phosphate-phosphate O...O distance 5.2 Å compared to 5.3-5.4 Å reported in the crystal structure of A-DNA[31]). The solvation shells of neighboring phosphate groups thus considerably overlap. The projection of electric field on the O=P=O bisector constitutes the relevant quantity of solvatochromic shifts of $\nu_{AS}(PO_2)^-$ modes.[15] Accordingly, different realizations of first solvation shell water molecules can lead to the component around 1245 cm$^{-1}$ via, e.g., the overlap of hydration spheres or a partial dehydration of phosphate groups due to steric boundary conditions imposed at the RNA surface.

The $\nu_{AS}(PO_2)^-$ component around 1280 cm$^{-1}$ (simulation: $\nu_{AS}(PO_2)^-$ = 1277.0 cm$^{-1}$, label C in Fig. 5a) arises from a phosphate group in contact with a counterion (panel C in Fig. 5c). One of the phosphate oxygen atoms replaces a water in the first solvation shell around the ion. Vice versa, the under-coordination of the phosphate group by water molecules imposes the substantial blue-shift of $\nu_{AS}(PO_2)^-$. Notably, the first solvation shell around the ion forms a hydrogen bond with a phosphate group on the opposite side of the narrow major groove.

Figure 5(b) presents hydrogen bond distributions for different phosphate group hydration structures of the RNA A-helix. For fully hydrated $(PO_2)^-$ groups (cf. panel A in Fig. 5c) we find that $(PO_2)^-$ oxygen atoms are coordinated on average by 2.41 water molecules, which agrees well with the hydration of B-DNA.[32] Partial under-coordination in the first solvation shell (cf. panels B,C in Fig. 5c) is mirrored in a reduced coordination of $(PO_2)^-$ oxygen atoms (2.01 water molecules on average) and the occurrence of bridging water molecules between adjacent phosphate groups (cf. Fig. 6). The partial hydration of phosphate groups in RNA reflects steric constraints imposed by the A-helix where phosphate oxygens point into the deep and narrow major groove, and the high electrostatic potential in the groove[33] yielding a high counterion population.

Hydration structures around phosphate groups of the RNA A-helix were further quantified via radial distribution functions g(r) of phosphate oxygen - water oxygen and phosphate oxygen – sodium ion (Na+) distances (Fig. 6). Compared to water oxygen-oxygen distances, phosphate oxygen-water oxygen distances are reduced (first maximum in g(r): 2.69 Å vs 2.78 Å[34]). Due to the negative charge of the phosphate group, the phosphate hydration shell is more rigid and structured. We observe characteristic differences in g(r) for the coordination of phosphate oxygen atoms characterized by under-coordination of first solvation shell water molecules (Fig. 5c, panels B,C) where the occurrence of bridging water molecules is reflected in a particular high value of g(r) at r ~ 3.2 Å (Fig. 6). Such bridging water hydration structures of phosphate group oxygen atoms closely mirror findings from X-ray crystallographic data.[7,31] Compared to the hydration shell of DNA where the phosphate oxygens are almost freely accessible by water molecules, the RNA phosphate oxygens point into the deep and narrow major groove. The water hydration structures are thus subject to higher steric constraints and the high electrostatic potential in the groove region is reflected in a higher ion population (Fig. 6).

Our experimental and theoretical results identify three distinct arrangements of hydrating water molecules around phosphate groups in the RNA backbone. Their abundance and, thus, relative strengths in the vibrational spectra change strongly with temperature. Around the dsRNA A-helix at $T_S \approx 300$ K, the two main hydration geometries are ordered water structures leading to a frequency $\nu_{AS}(PO_2)^- \approx 1245$ cm$^{-1}$ (Fig. 5c, panel B), and - to lesser extent - separate local phosphate hydration shells giving rise to $\nu_{AS}(PO_2)^- \approx 1220$ cm$^{-1}$ (Fig. 5c, panel A). The latter are very similar to the hydration shells around $(PO_2)^-$ groups in DNA[32] where the larger separation between neighboring phosphate groups and the absence of 2'-OH groups on the ribose units excludes the formation of ordered water patterns. With rising temperature $T_S$, the dsRNA structure becomes increasingly disordered, favoring DNA-like local hydration shells that at temperatures $T_S > T_m$ strongly dominate. Upon heating, there is a loss of phosphate/K$^+$ contact ion pairs, i.e., less shielding of the repulsive Coulomb interactions between neighboring phosphate groups.

The hydrated structure of ssRNA is characterized by a larger degree of disorder of the hydration patterns, as is evident from both the linear infrared and 2D-IR spectra, displaying a stronger weight of the component around 1220 cm$^{-1}$. A spectral feature indicative of contact ion pairs is absent in the ssRNA spectra, a fact which may point to a weaker repulsive Coulomb interaction between phosphate groups due to their – on the average - larger separation compared to dsRNA. For both dsRNA and ssRNA, the distribution of hydration geometries remains unchanged on ultrafast time scales. The frequency fluctuation correlation function accounting for the inhomogeneous broadening of the 2D-IR line shapes sets a lower limit on the order of 10 ps for hydrogen bond lifetimes in the first hydration shell around the phosphate groups, in line with simulation work in the literature.[4]

The thermal change of hydration structure established here is connected with changes of the hydration free energy.[35,36] A transition from the ordered water structure to DNA-like local phosphate hydration shells is accompanied by an increase in the number of phosphate water hydrogen bonds (Fig. 5b) and, thus, a decrease of enthalpy of solvation on the order of $\Delta H$= -(20-40) kJ/mole (cf. SI), together with a concomitant decrease of entropy. The enthalpy and entropy changes partly compensate each other, similar to the enthalpy-entropy compensation in RNA self-association.[36,37] However, the enthalpy and entropy changes due to hydration are of the same order of magnitude as those due to self-association of nucleobases. Thus, changes of the hydration free energy are a non-negligible part of the free energy balance upon RNA self-association and melting.

## 4. Conclusions

In conclusion, 2D-IR spectroscopy of asymmetric (PO$_2$)$^-$ stretching vibrations $\nu_{AS}$(PO$_2$)$^-$ of the RNA and DNA backbones represents a most sensitive approach to discern local hydration geometries along (double)-helical and disordered structures of the biomolecules. Combining this method with an in-depth theoretical analysis of molecular structure and dynamics provides quantitative information on the hydration patterns and the underlying interactions. An extension to other backbone vibrations in biomolecular structures of higher complexity is currently underway.

**Acknowledgments**. This research has received funding from the European Research Council (ERC) under the European Union's Horizon 2020 research and innovation program (grant agreements No. 833365 and No. 802817). B. P. F. acknowledges support by the DFG within the Emmy-Noether Program (Grant No. FI 2034/1-1). We thank Janett Feickert for expert technical support as well as Heike Nikolenko, Leibniz-Institut für Molekulare Pharmakologie (FMP), Berlin, for the opportunity to measure circular dichroism spectra.

**Supporting information available**: Experimental methods and results, theoretical methods and simulations.

**Figures**

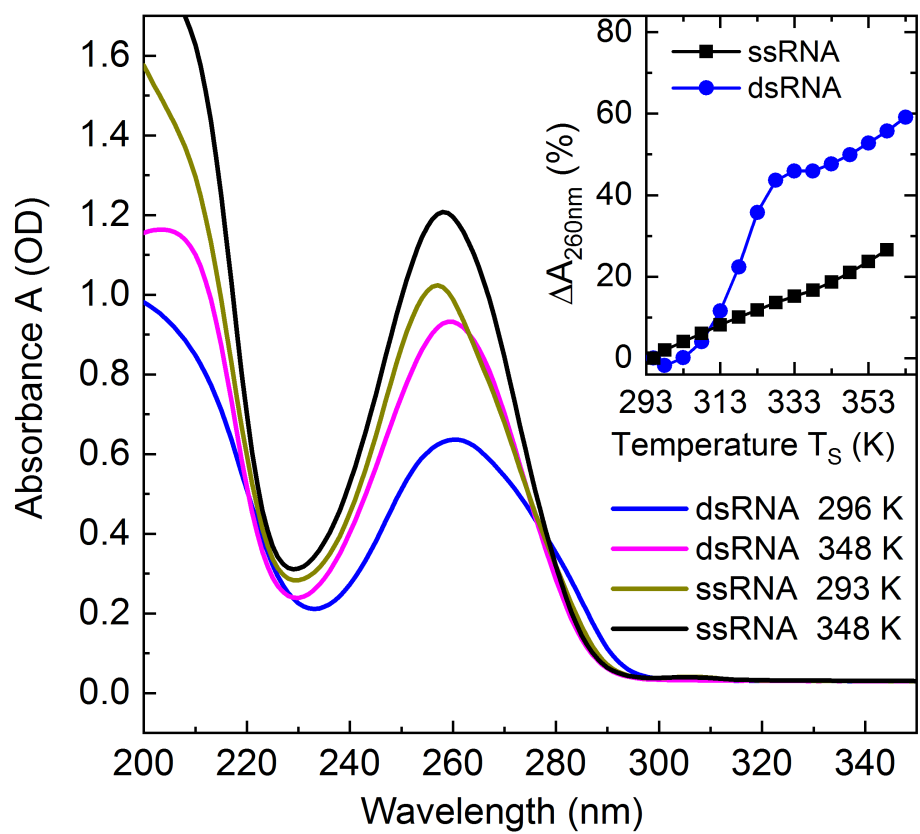

**Figure 1**. Ultraviolet absorption spectra for dsRNA and ssRNA in water at sample temperatures $T_S$=296 and 348 K. Inset: Differential peak absorbance $\Delta A_{260nm} = A_{260nm}(T_S) - A_{260nm}(296K)$ at 260 nm as a function of sample temperature $T_S$ ($A_{260nm}(T_S)$: absorbance at 260 nm at a sample temperature $T_S$).

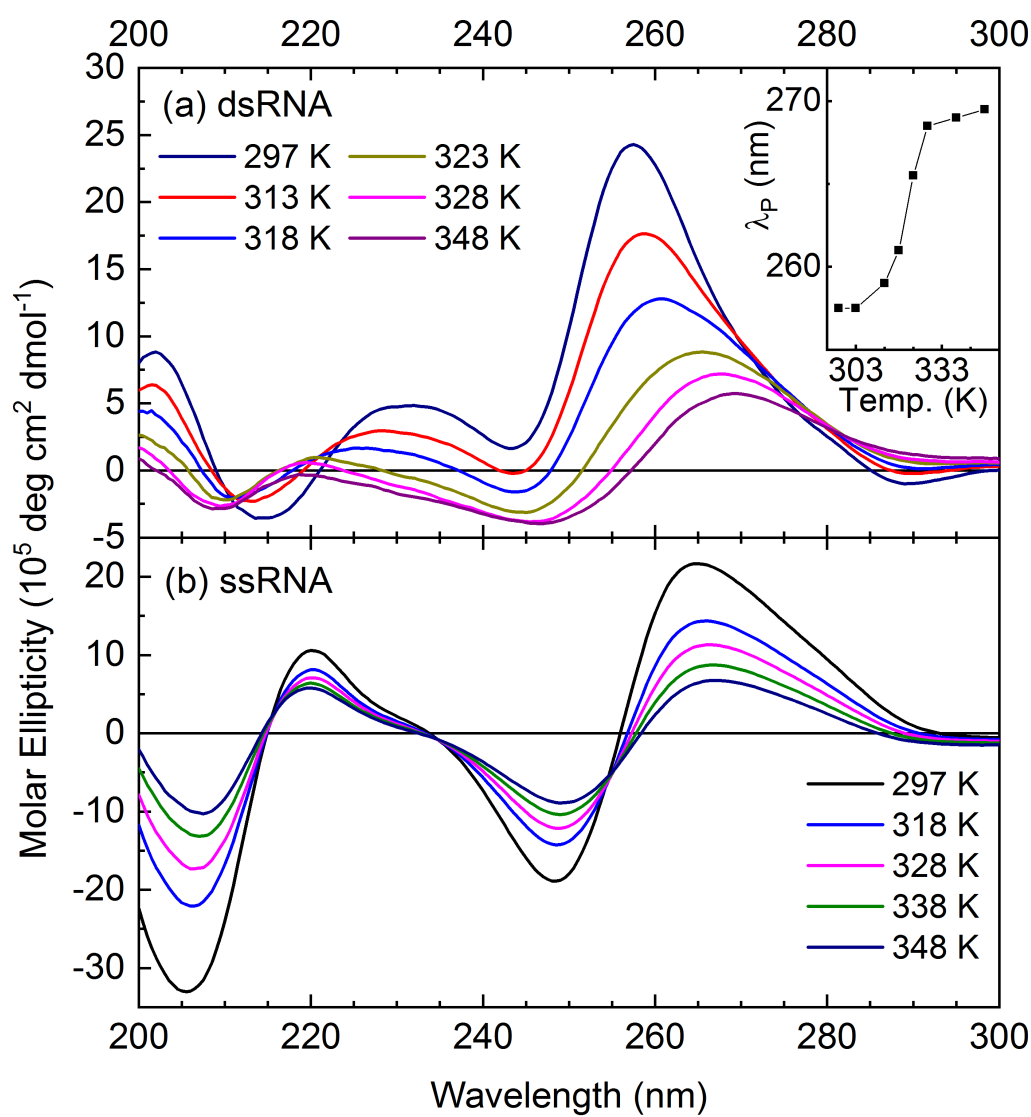

**Figure 2.** (a) CD spectra of dsRNA in water for different sample temperatures $T_S$ in the wavelength range from 200 to 300 nm. Inset: Temperature-dependent peak position $\lambda_P$ of the spectral maximum from 257.5 nm at $T_S$=298 K to 268.5 nm at $T_S$=348 K as derived from individual spectra. (b) CD spectra for ssRNA in water for different temperatures $T_S$.

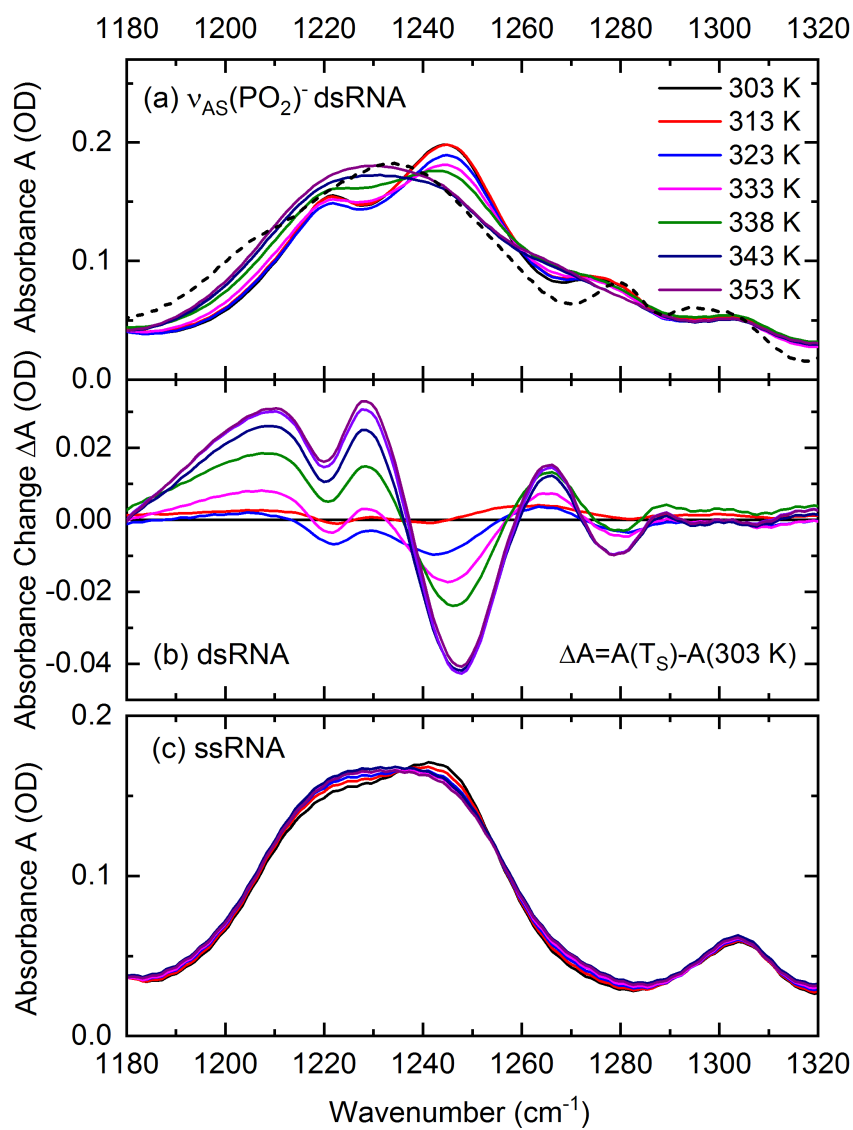

**Figure 3.** (a) Linear infrared absorption spectra of the asymmetric phosphate stretching vibration $\nu_{AS}(PO_2)^-$ of double-stranded RNA oligomers (dsRNA) in water for different sample temperatures $T_S$ (RNA concentration $c_{RNA}$=2.3 mM). Dashed line: infrared absorption of DNA oligomers at 303 K. This spectrum is normalized to the RNA peak absorption at 1230 cm$^{-1}$ (T=353 K). (b) Differential absorption spectra of dsRNA in the range of $\nu_{AS}(PO_2)^-$ for different temperatures. (c) Linear infrared absorption spectra of the asymmetric phosphate stretching vibration $\nu_{AS}(PO_2)^-$ of single-stranded RNA oligomers (ssRNA) in water for different sample temperatures $T_S$ (RNA concentration $c_{RNA}$=4.4 mM).

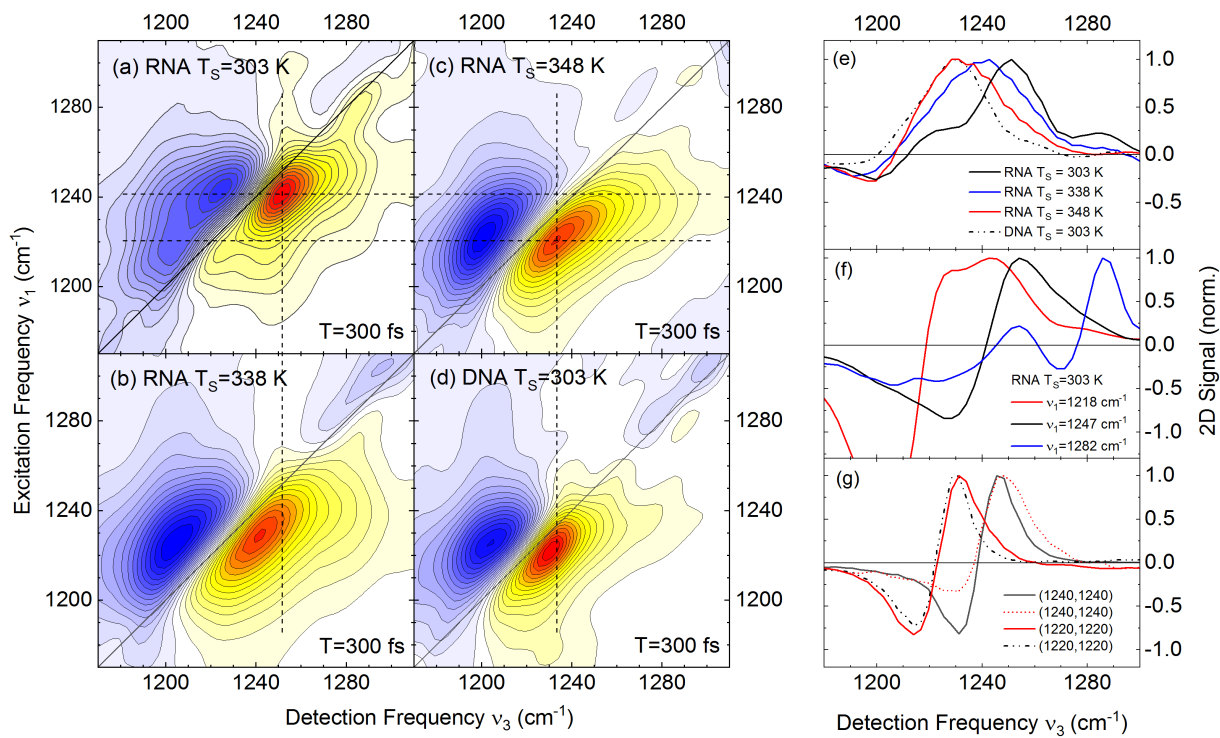

**Figure 4** (two-column). (a-d) 2D-IR spectra of the asymmetric phosphate stretching mode $\nu_{AS}(PO_2)^-$. (a) RNA oligomers at a sample temperature $T_S$=303 K, (b) RNA oligomers at $T_S$=338 K, (c) RNA oligomers at $T_S$=348 K, and (d) DNA oligomers at $T_S$=303 K. The absorptive 2D signal measured at a population time T=300 fs is plotted as a function of excitation frequency $\nu_1$ and detection frequency $\nu_3$. The yellow-red contours represent the signal due to the v=0→1 transition, the blue contours the signal on the v=1→2 transition. The signal change between neighboring contour lines is 7.5%. (e) Cuts of the 2D-IR spectra along a frequency diagonal crossing the position $(\nu_1,\nu_3)$=(1242,1250) cm$^{-1}$. The normalized 2D-IR signals are plotted as a function of detection frequency $\nu_3$. Data are shown for RNA at 3 different sample temperatures $T_S$ (solid lines) and DNA at $T_S$=303 K (dash-dotted lines). (f) Cuts of 2D-IR spectra along the detection frequency axis $\nu_3$ for the fixed excitation frequencies $\nu_1$ given in the inset. (g) Cuts of 2D-IR spectra of RNA and DNA along antidiagonals crossing the frequency positions given in the inset (in cm$^{-1}$). Black solid line: RNA at $T_S$=303 K, dash-dotted line: DNA at $T_S$=303 K. For RNA at $T_S$=348 K, cuts at (1240,1240) (red dotted line) and (1220,1220) cm$^{-1}$ (red solid line) are presented.

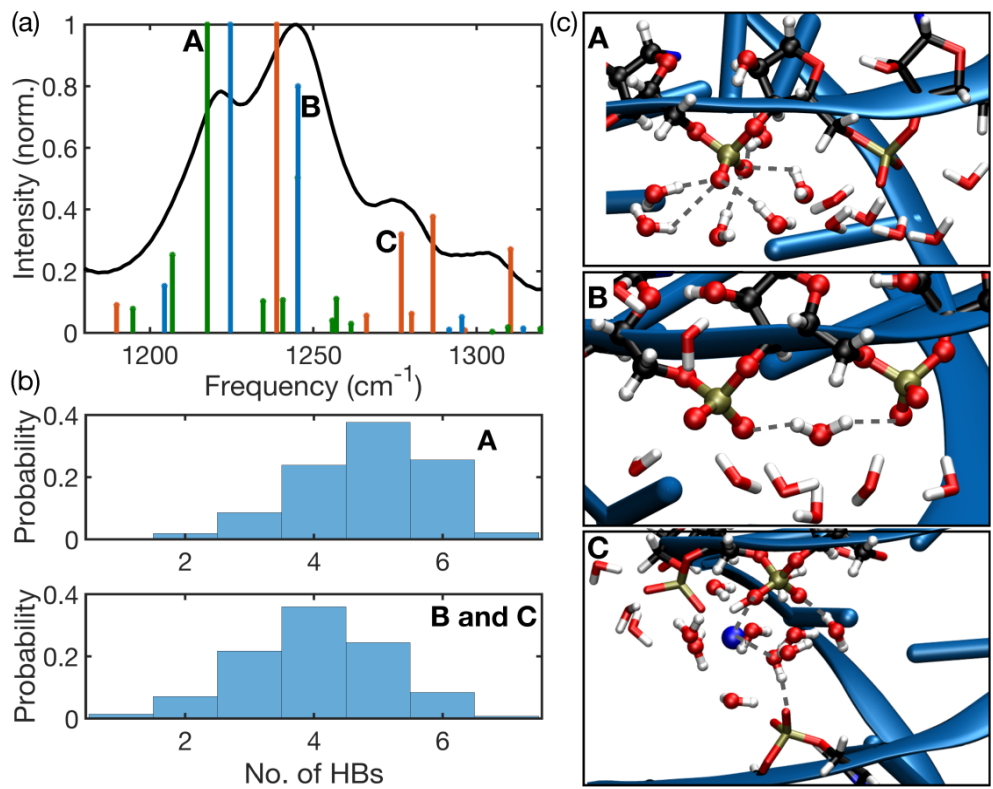

**Figure 5**. (a) Stick representation of simulated infrared absorption spectra of asymmetric stretching vibrations $\nu_{AS}(PO_2)^-$, normalized to the maximal intensity in the range 1180-1320 cm$^{-1}$. Green, blue and red colors denote vibrational frequencies obtained from QM/MM density functional simulations for three phosphate group solvation geometries A-C shown in Fig. 5 (c). Phosphate stretching vibrations labeled with A-C are located on the phosphate group highlighted in ball-and-stick representation together with particular first solvation shell waters and ions. The experimental infrared spectrum (black line) is shown for comparison. (b) Histogram of hydrogen bonds (HBs) in the first water layer around phosphate groups with different hydration structures (cf. panels A-C in Fig. 5c). Hydrogen bonds were averaged over six different 4-ns segments, equally separated in a 1.2 µs molecular dynamics trajectory. (c) Prototypical phosphate group solvation geometries A-C.

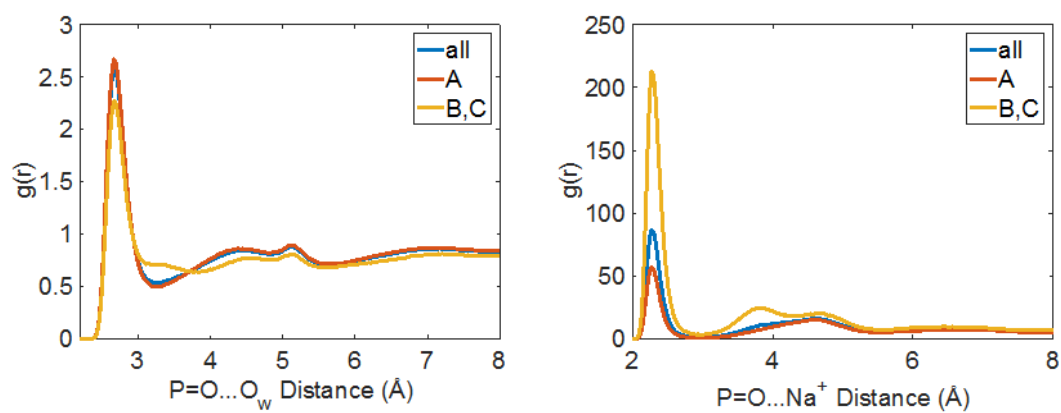

**Figure 6**. Radial distribution function g(r) of phosphate oxygen - water oxygen (P=O…Ow) distances (left) and phosphate oxygen – sodium ion (P=O…Na+) distances (right). Bridging water molecules of species B,C (left) are reflected in the high value of g(r) at r ~ 3.2 Å(7).

TOC Graphic

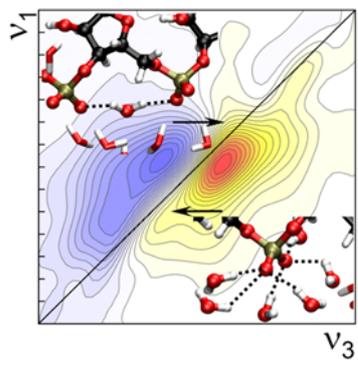